\documentclass[sigconf]{acmart}

\copyrightyear{2024} 
\acmYear{2024}
\setcopyright{acmlicensed}
\acmConference[ICSE '24]{2024 IEEE/ACM 46th International Conference on Software Engineering}{April 14--20, 2024}{Lisbon, Portugal}
\acmBooktitle{2024 IEEE/ACM 46th International Conference on Software Engineering (ICSE '24), April 14--20, 2024, Lisbon, Portugal}
\acmDOI{10.1145/3597503.3639142}
\acmISBN{979-8-4007-0217-4/24/04}

\usepackage{pifont}
\usepackage{multirow}
\usepackage[normalem]{ulem}
\usepackage{booktabs}
\usepackage{listings}
\usepackage[flushleft]{threeparttable}
\usepackage{balance}
\usepackage{url}
\usepackage{framed}

\newcommand{\mysubsection}[1]{\subsection{#1}}

\begin{document}

\title{Pre-training by Predicting Program Dependencies for Vulnerability Analysis Tasks}

\author{Zhongxin Liu}
\affiliation{
  \institution{The State Key Laboratory of Blockchain and Data Security, Zhejiang University}
  \country{China}
}
\email{liu_zx@zju.edu.cn}

\author{Zhijie Tang}
\affiliation{\institution{Zhejiang University}
\country{China}
}
\email{tangzhijie@zju.edu.cn}

\author{Junwei Zhang}
\affiliation{\institution{Zhejiang University}
\country{China}
}
\email{jw.zhang@zju.edu.cn}

\author{Xin Xia}
\authornote{Corresponding author.}
\affiliation{
  \institution{Huawei}
  \country{China}
}
\email{xin.xia@acm.org}

\author{Xiaohu Yang}
\affiliation{
  \institution{The State Key Laboratory of Blockchain and Data Security, Zhejiang University}
  \country{China}
}
\email{yangxh@zju.edu.cn}

\begin{abstract}
Vulnerability analysis is crucial for software security.
Inspired by the success of pre-trained models on software engineering tasks, this work focuses on using pre-training techniques to enhance the understanding of vulnerable code and boost vulnerability analysis.
The code understanding ability of a pre-trained model is highly related to its pre-training objectives.
The semantic structure, e.g., control and data dependencies, of code is important for vulnerability analysis.
However, existing pre-training objectives either ignore such structure or focus on learning to use it.
The feasibility and benefits of learning the knowledge of analyzing semantic structure have not been investigated.
To this end, this work proposes two novel pre-training objectives, namely Control Dependency Prediction (CDP) and Data Dependency Prediction (DDP), which aim to predict the statement-level control dependencies and token-level data dependencies, respectively, in a code snippet only based on its source code.
During pre-training, CDP and DDP can guide the model to learn the knowledge required for analyzing fine-grained dependencies in code.
After pre-training, the pre-trained model can boost the understanding of vulnerable code during fine-tuning and can directly be used to perform dependence analysis for both partial and complete functions.
To demonstrate the benefits of our pre-training objectives, we pre-train a Transformer model named PDBERT with CDP and DDP, fine-tune it on three vulnerability analysis tasks, i.e., vulnerability detection, vulnerability classification, and vulnerability assessment, and also evaluate it on program dependence analysis.
Experimental results show that PDBERT benefits from CDP and DDP, leading to state-of-the-art performance on the three downstream tasks.
Also, PDBERT achieves F1-scores of over 99\% and 94\% for predicting control and data dependencies, respectively, in partial and complete functions.

\end{abstract} 



\maketitle

\section{Introduction}~\label{sec:intro}

Software vulnerabilities are flaws or weaknesses that could be exploited to violate security policies~\cite{shirey2000internet}.
It is crucial to detect, categorize and assess vulnerabilities.
Due to the rapid increase in the number of software vulnerabilities and the success of deep learning techniques, researchers have proposed diverse deep-learning-based approaches to automate vulnerability analysis, such as 
vulnerability detection~\cite{zhou2019devign, chakraborty2022reveal}, classification~\cite{aota2020automation,zou2019mu}, patch identification~\cite{zhou2021finding,zhou2021spia} and assessment~\cite{le2021deepcva,le2023survey}, and achieved promising results.

Recently, inspired by the impressive effectiveness of pre-training large models in the natural language processing (NLP) field~\cite{devlin2019bert,brown2020languagea, raffel2020exploringa}, researchers proposed to pre-train large models on large-scale code-related corpora for capturing the common knowledge of programming languages.
We refer to such models as \textit{pre-trained code models}.
Pre-trained code models have shown consistent performance improvements over the neural models without being pre-trained (for short, \textit{non-pre-trained models}) on various software engineering (SE) tasks, e.g., code search~\cite{feng2020codebert}, code clone detection~\cite{guo2021graphcodebert}, and code completion~\cite{svyatkovskiy2020intellicodea}.
Motivated by these studies, this work aims to boost vulnerability analysis using pre-training, with a focus on helping neural models better understand vulnerable code through pre-training techniques.

The code understanding ability of a pre-trained code model largely hinges on its pre-training objectives.
Effective objectives can guide the model to learn the prior knowledge that is helpful for downstream tasks.
Early pre-trained code models~\cite{feng2020codebert,kanade2020learninga} directly use the pre-training objectives designed for natural languages (NL), e.g., Masked Language Model (MLM)~\cite{devlin2019bert}, ignoring the syntactic and semantic structure of code.
Recently, some researchers~\cite{wang2022codemvp,ding2022learning} specifically design several pre-training tasks, e.g., predicting Abstract Syntax Tree (AST) node types, to capture the syntactic structure of code.

The semantic structure of code, e.g., control and data dependencies~\cite{ferranteProgramDependenceGraph1987}, plays an important role in vulnerability analysis~\cite{zhou2019devign,chakraborty2022reveal,cao2022mvd}.
For example, to detect a use-after-free vulnerability, a model needs to identify whether the argument of a memory-releasing function call is used (which relies on data dependencies) in a statement that may be executed after this call (which is related to control dependencies).
To the best of our knowledge, only two pre-trained code models, i.e., \textsc{Code-MVP} and GraphCodeBERT, consider the semantic structure of code.
\textsc{Code-MVP} takes control flow graph (CFG) as input during pre-training.
GraphCodeBERT takes as input data flow edges during pre-training and inference.
On the other hand, prior studies have proposed some non-pre-trained models that consider semantic structure for vulnerability analysis~\cite{li2018vuldeepecker,zhou2019devign,chakraborty2022reveal}.
These approaches first extract control and data dependencies, i.e., program dependencies, using static analysis tools, e.g., Joern~\cite{yamaguchi2014modeling}, and then train a neural model to learn from the extracted dependencies.
These pre-trained and non-pre-trained models have two main limitations.
First, they require the input code to be correctly parsed to extract its semantic structure, and cannot handle partial code (e.g., incomplete functions).
However, the ability to analyze partial code is useful and valuable for some real-world scenarios, e.g., detecting vulnerable code snippets shared on Stack Overflow~\cite{verdi2020empirical}.
Second, they target learning representation \textbf{from} the extracted semantic structure, or learning to use such structure.
Considering that different downstream tasks can utilize the semantic structure in various ways, the generality of such representation can be limited.

To this end, this work introduces the idea of pre-training code models to incorporate the knowledge required for end-to-end program dependence analysis (i.e., from source code to program dependencies), and proposes two novel pre-training objectives, i.e., Control Dependency Prediction (CDP) and Data Dependency Prediction (DDP), to instantiate this idea.
Specifically, CDP and DDP ask the model to predict all statement-level control dependencies and token-level data dependencies in a program relying solely on source code.
Different from existing pre-training objectives, CDP and DDP target guiding the model to learn the representation that \textbf{encodes} the semantic structure of code \textbf{only based on source code}.
However, it is not easy to train DDP.
Because the number of token-level data dependencies is much lower than all possible token pairs in code, i.e., DDP suffers from severe data imbalance.
To address this problem, we leverage a masking strategy to filter impossible token pairs and enable DDP to function effectively.

To demonstrate the effectiveness of CDP and DDP, we pre-trained a Transformer model named PDBERT (\textbf{P}rogram \textbf{D}ependence \textbf{BERT}) on 1.9M C/C++ functions using CDP and DDP as well as MLM.
CDP and DDP bring several benefits to PDBERT:
First, PDBERT only requires source code as input and is capable of handling the code snippets that cannot be correctly parsed, e.g., partial code.
Second, PDBERT incorporates the knowledge of end-to-end program dependence analysis, which is more general than the knowledge of using program dependencies, and can better boost diverse downstream tasks.
Third, PDBERT can be directly used to perform program dependence analysis for partial and complete code.
Please note that existing static analysis tools, e.g., Joern~\cite{yamaguchi2014modeling}, cannot correctly derive program dependencies in partial code without manual intervention.
Thus, although program dependence analysis is only one of PDBERT's usage scenarios, PDBERT complements static analysis tools with the ability to analyze partial code.

We evaluate PDBERT in both intrinsic and extrinsic ways.
For intrinsic evaluation, we directly apply PDBERT to analyze control and data dependencies for both partial and complete functions based only on their source code.
Experimental results show that the F1-scores of PDBERT for identifying statement-level control dependencies and token-level data dependencies are over 99\% and 95\%, respectively, for partial functions, and over 99\% and 94\% for complete functions.
These results indicate that PDBERT successfully learns the knowledge of program dependence and can effectively analyze both partial and complete functions.
Moreover, the throughput of PDBERT is \textbf{23 times} higher than the state-of-the-art program dependence analysis tool Joern, indicating that PDBERT is more suitable for the use cases where some low levels of imprecision are tolerant and the throughput matters more.
For extrinsic evaluation, we fine-tune and evaluate PDBERT on three vulnerability analysis tasks, i.e., vulnerability detection, vulnerability classification, and vulnerability assessment.
PDBERT benefits from CDP and DDP and outperforms the best-performing baselines by 5.9\%-9.0\% on the three tasks.

In summary, the contributions of this work are as follows:
\begin{itemize}
    \item We introduce the idea of pre-training code models to incorporate the knowledge required for end-to-end program dependence analysis, and propose two novel pre-training objectives, i.e., CDP and DDP, to instantiate this idea.
\item We have built a pre-trained model named PDBERT with CDP, DDP and MLM, which to the best of our knowledge, is the first neural model that can analyze statement-level control dependencies and token-level data dependencies.
\item We conduct both intrinsic and extrinsic evaluations, which show that PDBERT can accurately and efficiently identify program dependencies in both partial and complete functions, and can facilitate diverse vulnerability analysis tasks.
\end{itemize}

Our replication package is available at~\cite{website:github,website:replicate}.

 \section{Preliminary}\label{sec:background}
This section introduces Program Dependence Graph (PDG) and describes why handling partial code is useful.

\begin{figure}[t]
    \centering
    \includegraphics[width=0.44\textwidth]{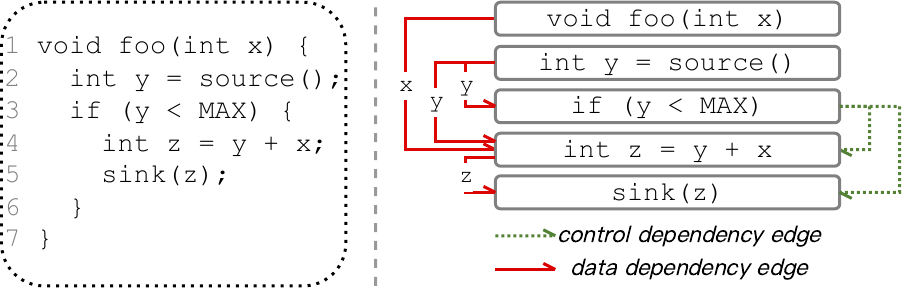}
\caption{A sample function and its PDG.}
\label{fig:pdg}
\end{figure}
     
\mysubsection{Program Dependence Graph}\label{sec:pdg}
Program dependencies, including control and data dependencies, reflect the semantic structure of code.
Program dependence graph (PDG)~\cite{ferranteProgramDependenceGraph1987} explicitly represents control and data dependencies with different types of edges and is frequently adopted by previous studies for vulnerability analysis\cite{zhou2019devign,chakraborty2022reveal,cao2022mvd}. 
Figure~\ref{fig:pdg} represents a sample function and its PDG.
In a PDG, each node denotes a statement or a predicate, and there are two types of edges: control dependency edges and data dependency edges.
Control dependency edges (dotted green lines in Figure~\ref{fig:pdg}) present the control flow relationships between predicates and statements, and can be used to infer the control conditions on which a statement depends.
Data dependency edges (solid red lines in Figure~\ref{fig:pdg}) present the def-use relationships, and each of them is labeled with a variable that is defined in the source node and used in the target node.
By predicting control and data dependencies in programs, the model can learn to strengthen the connections between computationally related parts of a program and better capture the semantic structure of code.

\mysubsection{Usage Scenario of Handling Partial Code}
Compared to prior work, one advantage of PDBERT is that it can handle partial code, including performing program dependence analysis and downstream tasks on partial code.
One usage scenario of this feature is analyzing the code snippets on Stack Overflow.
Prior work~\cite{acar2016you, abdalkareem2017on} has shown that novices and even more senior developers copy code snippets from Stack Overflow into production software.
If the copied snippets are vulnerable, their production software will be prone to attacks.
Specifically, Fischer et al.~\cite{fischer2017stack} observed that insecure code snippets from StackOverflow are copied into popular Android applications installed by millions of users.
From 72K C++ code snippets used in at least one GitHub repository, Verdi et al.~\cite{verdi2020empirical} found 99 vulnerable code snippets of 31 types, and these snippets had affected 2,859 GitHub projects.
Considering these facts, it is essential and valuable to analyze the code snippets on Stack Overflow before they are scattered to other places.
We argue that PDBERT can be used as a fundamental tool/model in this scenario.

 \section{Approach}\label{sec:approach}
This section first describes how PDBERT represents the input (\S\ref{sec:input}) and how to construct ground truth for CDP and DDP (\S\ref{sec:pde} and \S\ref{sec:ddc}).
Then, we elaborate on the three pre-training tasks used by PDBERT (\S\ref{sec:pre_training_tasks}) and the usages of PDBERT (\S\ref{sec:usage}).

\mysubsection{Input Representation}\label{sec:input}
PDBERT only requires source code as input.
Given the source code $C$ of a program, PDBERT first uses a subword-based tokenizer, e.g., a Byte-Pair Encoding (BPE) tokenizer~\cite{senrich2016neural}, to tokenize it into a sequence of tokens and prepends this sequence with a special token [CLS].
The resulting token sequence, denoted as $T=[t_{cls},t_1,...,t_{n}]$, is then fed into a multi-layer Transformer model to obtain the contextual embedding of each token.
We denote these contextual embeddings as $H^t=[h^t_{cls},h^t_1,...,h^t_{n}]$.
For each token $t_i$, its char span is represented as $S_{t_i} = [I_{t_i}^l, I_{t_i}^r]$ and recorded, where $I_{t_i}^l$ and $I_{t_i}^r$ denote the start and end character indices of $t_i$ in the source code.

\mysubsection{Program Dependency Extraction}\label{sec:pde}

To construct ground truth for CDP and DDP, we need to extract control and data dependencies in programs.
Given a program, we first leverage a static analysis tool named Joern~\cite{yamaguchi2014modeling} to construct its PDG and AST based on its source code, and combine them into a joint graph.
Then, we extract control and data dependencies from this graph.

\subsubsection{Control Dependency Extraction.}\label{sec:cde}
As described in Section~\ref{sec:pdg}, the control dependency edges in the PDG represent control dependencies.
We encode them at the statement level into a matrix $G^c \in \mathcal{R}^{m^c \times m^c}$, where $m^c$ denotes the maximum number of nodes in the PDG and is set to 50 by default.
$G^c_{i,j} \in \{1, 0\}$ denotes whether the $j$-th PDG node (statement) is control-dependent on the $i$-th PDG node (predicate).

\subsubsection{Data Dependency Extraction.}\label{sec:dde}
Similar to control dependencies, statement-level data dependencies can be extracted based on the data dependency edges in the PDG and encoded as a matrix.
However, data dependencies are naturally defined on variables.
Such matrix is coarse-grained, ignoring the information of variables.
Therefore, we extract and encode data dependencies at a finer level to consider variable information and capture finer-grained code structure.
Recall that each data dependency is represented by a data dependency edge associated with a variable \texttt{var} in the PDG.
For the $i$-th data dependency edge $D_i$, we first identify the two AST nodes that correspond to the \texttt{var} defined in $D_i$'s source PDG node and the \texttt{var} used in $D_i$'s target PDG node from the joint graph.
We refer to them as the \textit{start node} $s_i$ and the \textit{end node} $e_i$ of $D_i$, respectively.
Then, we represent $D_i$ as the combination of $s_i$ and $e_i$, i.e., $D_i = [s_i, e_i]$.
Consequently, the data dependencies in the program are represented as $\hat{G}^d = [D_1, D_2, ...]$.

\begin{figure}[t]
    \includegraphics[width=0.49\textwidth]{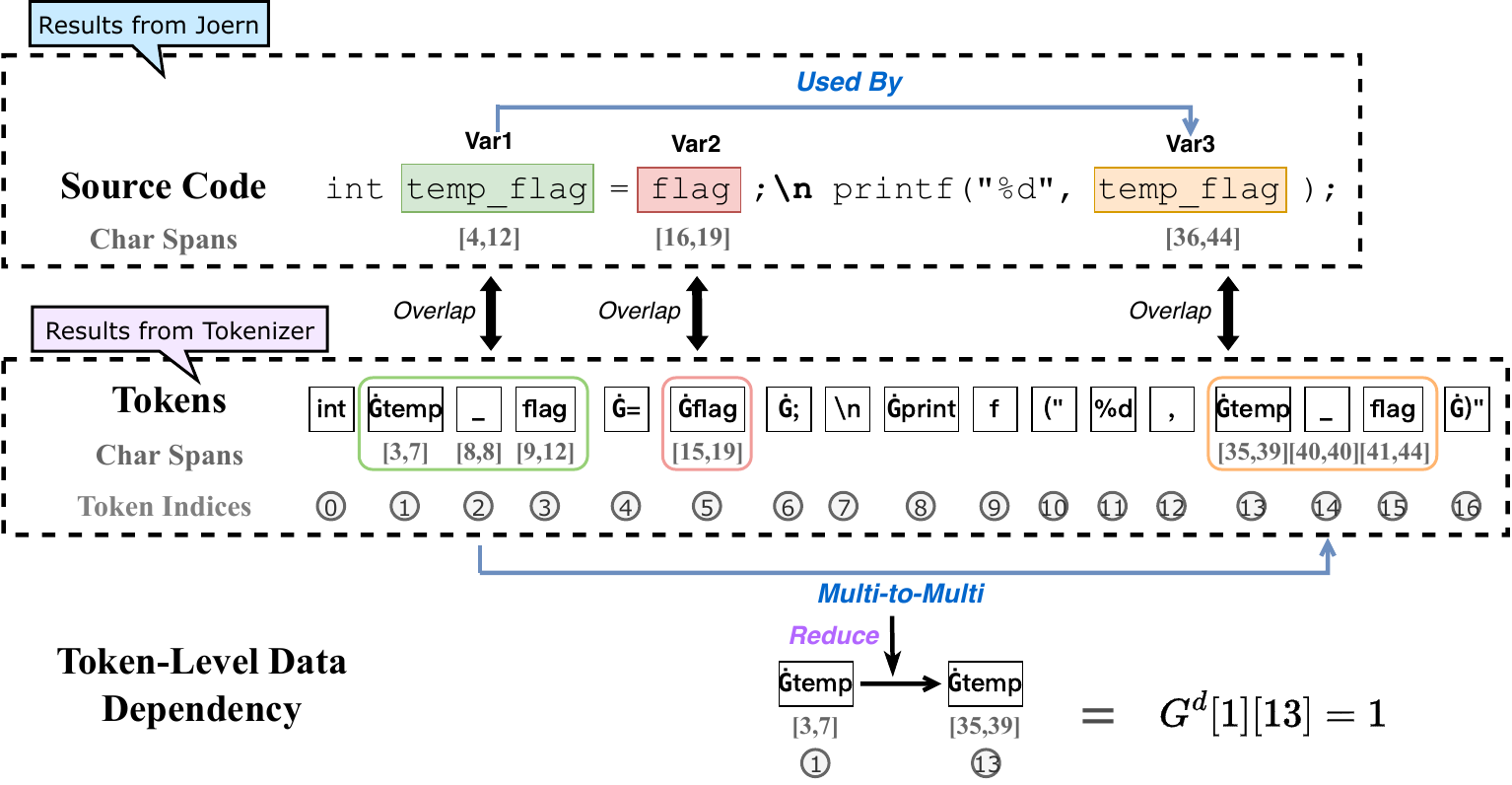}
\caption{Constructing the ground truth $G^d$ of DDG.}  
    \label{fig:char_span_intersection}
\end{figure}

\mysubsection{Token-Level Data Dependency Construction}\label{sec:ddc}
DDP aims to produce token-level data dependencies in the input program.
In Section~\ref{sec:pde}, data dependencies have been encoded as a set of AST node pairs $\hat{G}^d$.
Each AST node corresponds to a code element.
But each code element may be mapped into multiple tokens in $T$ due to the subword-based tokenizer.
Therefore, we need to refine $\hat{G}^d$ to represent token-level data dependencies.
As illustrated by Figure~\ref{fig:char_span_intersection}, we achieve this by converting $\hat{G}^d$ into a matrix $G^d$, where $G^d_{i,j} \in \{0, 1\}$ denotes whether the $j$-th token is data-dependent on the $i$-th token in $T$, as follows:

\textbf{Mapping AST nodes to tokens}. We retrieve the code element that corresponds to each AST node $a_i$ in $\hat{G}^d$, and map the code element into its char span $S_{a_i}=[I^{l}_{a_i}, I^{r}_{a_i}]$, where $I^{l}_{a_i}$ and $I^{r}_{a_i}$ denote the indices of $a_i$'s start and end characters in the source code, respectively.
For example, in Figure~\ref{fig:char_span_intersection}, the char span of the first \texttt{temp\_flag} is [4,12].
Based on the char spans of $a_i$ and the tokens in $T$, we map $a_i$ into a subsequence of $T$, namely $T_{a_i}$.
Specifically, we find the tokens in $T$ of which the char spans are overlapped with $a_i$'s char span, and gather these tokens in order to construct $T_{a_i}=[t^{a_i}_1,t^{a_i}_2,...]$.
For example, in Figure~\ref{fig:char_span_intersection}, each \texttt{temp\_flag} is mapped to [\texttt{Ġtemp}, \texttt{\_}, \texttt{flag}], where Ġ is used by the tokenizer to denote the space before a word.
After this mapping, each data dependency edge $D_i$ is represented as a pair of token sequences $T_{s_i}$ and $T_{e_i}$, where $s_i$ and $e_i$ are the start and end nodes of $D_i$.

\textbf{Generating token edges}. Based on the new representations of data dependency edges, each token-level data dependency can be represented by the edges between the tokens in $T_{s_i}$ and the tokens in $T_{e_i}$.
Intuitively, we can connect the tokens in $T_{s_i}$ to the tokens in $T_{e_i}$ using the Cartesian production.
However, this will produce plenty of edges and most of them are unnecessary.
Therefore, for each data dependency edge $D_i$, we choose to connect only the first tokens in $T_{s_i}$ and $T_{e_i}$, which effectively represents this data dependency and produces only one token-level edge, i.e., $t^{s_i}_1 \to t^{e_i}_1$.
For example, in Figure~\ref{fig:char_span_intersection}, the first tokens of the two \texttt{temp\_flag} are connected.
After this step, each data dependency edge $D_i$ is represented as a pair of tokens, e.g., $D_i = [t^{s_i}_1, t^{e_i}_1]$.

\textbf{Constructing $G^d$}. We initialize $G^d \in \mathcal{R}^{m^d \times m^d}$ as a matrix filled with zeros, where $m^d$ is the max sequence length of the model.
Each element in $G^d$ represents the data dependency between two tokens in the code.
For each data dependency edge $D_i = [t^{e_i}_1, t^{e_i}_1]$, we retrieve the token indices of $t^{s_i}_1$ and $t^{e_i}_1$ in $T$, assuming that they are $x$ and $y$, and set $G^d_{x,y}$ to 1.
For the data dependency edge presented in Figure~\ref{fig:char_span_intersection}, $x=1$, $y=13$, so $G^d_{1,13}$ is set to 1.
$G^d$ will be used as the ground truth of DDP.

\begin{figure*}[!t]
    \includegraphics[width=0.92\textwidth]{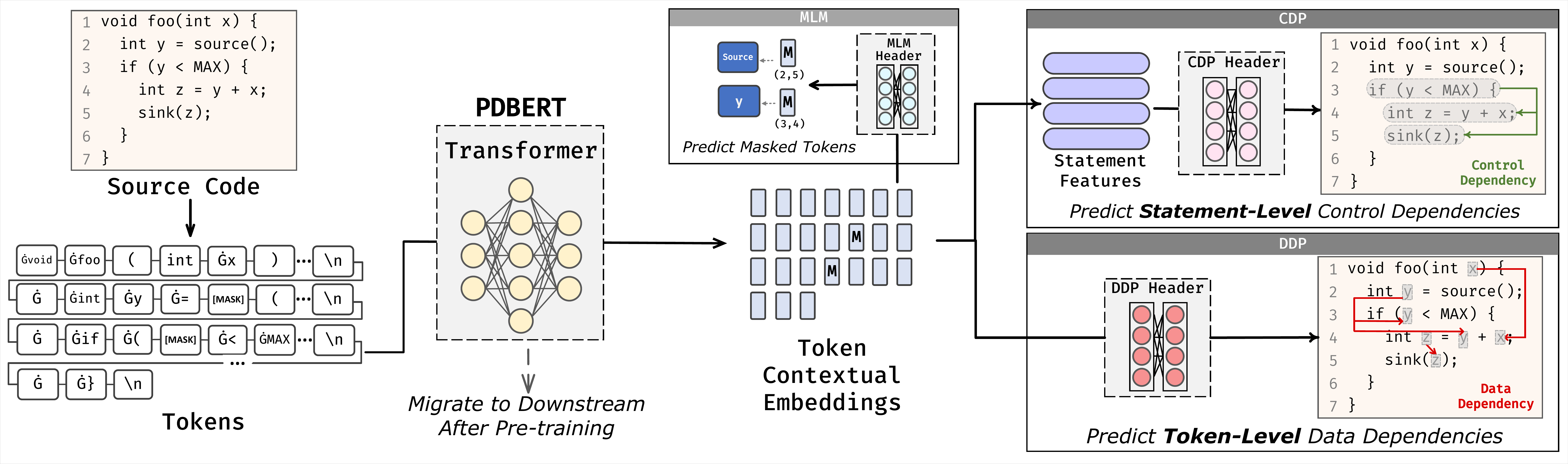}
\caption{The pre-training tasks of PDBERT.}
\label{fig:query-back}
\end{figure*} 
\mysubsection{Pre-training Tasks}\label{sec:pre_training_tasks}
PDBERT is pre-trained using three tasks: Masked Language Model (MLM), statement-level Control Dependency Prediction (CDP) and token-level Data Dependency Prediction (DDP), as shown in Figure~\ref{fig:query-back}.
MLM is widely used by existing pre-trained models \cite{liu2019roberta,feng2020codebert,guo2021graphcodebert}.
When used for programming languages (PL), it can guide the model to capture the naturalness and to some extent the syntactic structure of code~\cite{wan2022what}.
CDP and DDP can help the model learn the knowledge about the semantic structure of code, strengthening the attention between computationally related code elements.
The three tasks complement each other.

\subsubsection{Masked Language Model (MLM)}
MLM requires the model to reconstruct randomly masked tokens from the corrupted input sequence.
Before being encoded by the Transformer model, a certain proportion of tokens in $T$ are sampled and replaced with a special token \texttt{[MASK]}.
Based on the contextual embeddings $H^t$ produced by the Transformer model, we use a two-layer MLP (Multi-Layer Perceptions) with the Softmax function to predict the original token of each \texttt{[MASK]}, and calculate the loss of MLM, as follows:
\begin{equation}
    P(t_i|h^t_i) = \text{Softmax}_{t_i \in V}(\text{MLP}(h^t_i))
\end{equation}
\vspace{-0.2cm}
\begin{equation}
    \mathcal{L}^{mlm}= \frac{1}{|M^t|}\sum_{i \in M^t}-\log{P(t_i|h^t_i)}
\end{equation}
where V is the vocabulary of the model, Softmax$_{t_i \in V}$ denotes the probability of producing $t_i$, and $M^t$ are the indices of the sampled tokens.
Following prior work~\cite{liu2019roberta,feng2020codebert,guo2021graphcodebert}, we sample 15\% of tokens to mask.
We also replace 10\% of them with random tokens and unmask another 10\% of them to alleviate the inconsistency between pre-training and fine-tuning.

\subsubsection{Statement-Level Control Dependency Prediction (CDP)}
\label{statement level cdg prediction}
CDP aims to predict all the statement-level control dependencies in a program based on its source code.
As described in Section~\ref{sec:cde}, we encode the ground truth of CDP as a matrix $G^c$.
To predict $G^c$, 
we encode each PDG node into a feature vector based on the contextual embeddings $H^t$, and predict whether there is a control dependency between each pair of PDG nodes based on their feature vectors.
In detail, first, for each PDG node $q_k$ we identify the tokens in $T$ that belong to $q_k$ and average the contextual embeddings of these tokens to obtain a vector $\hat{h}^q_k$.
Next, we input $\hat{h}^q_k$ into a one-layer MLP activated by the ReLU function for dimension reduction and obtain a vector $h^q_k$, as follows:
\begin{equation}
    \hat{h}^q_k=\frac{1}{|q_k|}\sum_{t_i \in q_k}h^t_i
\end{equation}
\vspace{-0.2cm}
\begin{equation}
    h^q_k=\text{ReLU}(\text{MLP}(\hat{h}^q_k))
\end{equation}
where $|q_k|$ is the number of tokens belonging to $q_k$.
$h^q_k$ is regarded as the feature vector of $q_k$.
Then, we use a bilinear layer, a commonly used module for relation predictions \cite{socher2013reasoning}, to predict the probability $P^c_{i,j}$ that a PDG node $q_j$ is control-dependent on another PDG node $q_i$, as follows:
\begin{equation}
    P^c_{i,j} = \sigma(h^{q\top}_{i} W_c h^q_j + b_c)
\end{equation}
where $W_c$ and $b_c$ are trainable parameters and $\sigma$ is the Sigmoid function. 
Because the inputs of the bilinear layer are not commutative, the predicted control dependencies between two PDG nodes are asymmetric, i.e., $P^c_{i,j}$ is not necessarily equal to $P^c_{j,i}$.
This is in accord with the fact that control dependencies are directional. 
Finally, we calculate the cross entropy as the loss of CDP based on the predicted probabilities and the ground truth $G^c$ of size $|G^c|$:
\begin{equation}
    \mathcal{L}^{cdp}=\frac{1}{|G^c|}\sum_{i,j}-G^c_{i,j}\log{P^c_{i,j}} - (1-G^c_{i,j})\log{(1-P^c_{i,j})}
\end{equation}

\subsubsection{Token-Level Data Dependency Prediction (DDP)}
\label{sec:ddp}
DDP targets predicting all the token-level data dependencies in a program only based on its source code.
As described in Section~\ref{sec:ddc}, the ground truth of DDP is encoded as a matrix $G^d$.
The contextual embeddings $H^t$ are naturally the feature vectors of the tokens in $T$.
To predict $G^d$, DDP also uses a one-layer MLP with the ReLU function for dimension reduction and a bilinear layer for prediction, as follows:
\begin{equation}
    \hat{h}^t_i = \text{ReLU}(\text{MLP}(h^t_i))
\end{equation}
\begin{equation}
    P^d_{i,j} = \sigma(\hat{h}^{t\top}_i W_d \hat{h}^t_j + b_d)
\end{equation}
where $W_d$ and $b_d$ are trainable parameters. 
Nevertheless, compared to $G^c$, $G^d$ focuses on token-level relations and there are usually a large number of elements in $G^d$.
For example, if the length of $T$ is 512, the size of $G^d$ would be over 262,144.
Even worse, considering that the number of data dependencies is much lower than all possible token pairs in a program, $G^d$ is usually highly sparse.
If DDP directly predicts all elements in $G^d$, the model will mostly learn from the large proportion of zeros in $G^d$ and constantly predict that there is no data dependency between two tokens.

To alleviate this problem, we propose a token-type-based masking strategy to 
mask the values associated with one or more non-Identifiers in $G^d$.
Specifically, we first use a lexer to tokenize the input program $C$ into a sequence of \textit{code tokens} $T'=[t'_1, t'_2, ..., t'_{|T'|}]$ with their token types (e.g., Keyword and Identifier) labeled and char spans recorded.
Note that $T'$ is different from the token sequence $T$ produced by the subword-based tokenizer, and each code token $t'_i$ may be overlapped with multiple tokens in $T$.
Next, for each code token $t'_i$, we identify its overlapping tokens in $T$ based on their char spans and gather such tokens in order into a token sequence $T_{t'_i}$.
The token type of $t'_i$ is propagated to each token in $T_{t'_i}$.
Then, we mask non-Identifier tokens in $T$ and calculate the loss of DDP as follows:
\vspace{-0.1cm}
\begin{equation}
\mathcal{L}^{ddp}=\frac{\sum_{i,j}m^d_i m^d_j[-G^d_{i,j}\log{P^d_{i,j}} - (1-G^d_{i,j})\log{(1-P^d_{i,j})}]}{(\sum^{n}_{k=1}m^d_k)^2}
\end{equation}
\vspace{-0.1cm}
$m^d_i \in \{0,1\}$ indicates whether $t_i$'s token type is Identifier.
This formula means that the prediction loss of $G^d_{i,j}$ will be masked if the token type of either $t_i$ or $t_j$ is not Identifier.
It is also possible to introduce more and stronger priors to tackle this problem, e.g., only predicting data dependencies between identical identifiers.
We only consider token types, which can help the model learn more knowledge related to program dependence analysis, e.g., the knowledge of recognizing identical identifiers.

Besides this masking strategy, another way to consider token types during pre-training is to explicitly provide identifier pairs to the model for prediction.
However, during inference, this way requires additional tools to gather identifier pairs, while the masking strategy can be easily disabled to enable end-to-end program dependence analysis.
Also, the masking strategy implicitly guides the model to focus on identifier pairs, which can help the model learn more syntactic knowledge of code.

\mysubsection{The Usages of PDBERT}\label{sec:usage}
PDBERT takes as input only a code snippet, which can be either a partial or a complete function.
Like existing pre-trained models~\cite{devlin2019bert,feng2020codebert,guo2021graphcodebert}, PDBERT outputs a sequence of contextual embeddings by itself.
Such embeddings are converted to task output, such as a classification label or a token sequence, by a header (i.e., an output layer).
During pre-training, different headers are connected to PDBERT for different pre-training tasks, as shown in Figure~\ref{fig:query-back}.
After pre-training, we can simply connect PDBERT with the headers of CDP and DDP to infer the statement-level control dependencies and the token-level data dependencies in a code snippet.
For a downstream task, a new header (usually an MLP) is connected to PDBERT and fine-tuned with PDBERT on a task-specific dataset.
Each header is task-specific and is trained to contain the information about producing task output based on the contextual embeddings output by PDBERT.
Such information is usually unhelpful for other tasks~\cite{chen2020simple}.
As the headers of MLM, CDP and DDP are task-specific and are not used for downstream tasks, we argue they have no side effects.
Please note that PDBERT only takes as input the source code and does not require parsing the input program or constructing its PDG during fine-tuning.
We hypothesize that the knowledge of program dependence has been learned and absorbed by PDBERT during pre-training, and such knowledge can boost the downstream tasks that are sensitive to program dependencies.

 \section{Pre-training Setup}\label{sec:exp_setup}
This section describes the dataset and the configurations used to pre-train our model.

\mysubsection{Pre-training Dataset}~\label{sec:dataset}
\label{pre-training dataset}
In this work, we target C/C++ vulnerabilities.
We use the dataset collected by Hanzi et al.~\cite{hanzif2022vulberta} as our pre-training dataset.
It contains over 2.28M C/C++ functions either extracted from the top 1060 C/C++ open-source projects on GitHub or gathered from the Draper dataset~\cite{russel2019automated}.
We shuffle these functions and partition them into the training, validation and test sets, referred to as $PT_{train}$, $PT_{val}$ and $PT_{test}$, respectively.
$PT_{val}$ is used for hyperparameter tuning during pre-training.
$PT_{test}$ is leveraged to evaluate PDBERT on program dependence analysis.
To construct ground truth for CDP and DDP, we leverage Joern to build the AST and PDG of each function. 
Some functions in this dataset cannot be analyzed by Joern in a reasonable time, and are filtered by setting timeout to 30 minutes. 
In addition, we deduplicate $PT_{valid}$ and $PT_{test}$ and remove the functions in them that also appear in $PT_{train}$
Finally, $PT_{train}$, $PT_{val}$ and $PT_{test}$ contain about 1.9M, 155.0K and 60.4K C/C++ functions, respectively.

\mysubsection{Pre-training Model Configurations}
Following CodeBERT~\cite{feng2020codebert}, PDBERT uses a 12-layer Transformer with 768 dimensional hidden states and sets the max sequence length to 512.
During pre-training and fine-tuning, we truncate the code snippets with more than 512 tokens or $m^c$ statements to fit the model.
CDP and DDP only consider the control and data dependencies within the tokens input to the model.
We initialize PDBERT with the parameters of CodeBERT to accelerate the training process following GraphCodeBERT~\cite{guo2021graphcodebert}.
The pre-training objective of PDBERT is to jointly minimize the losses of the three pre-training tasks:
\vspace{-0.1cm}
\begin{equation}
\label{eq:loss}
    \mathcal{L}=a_1\mathcal{L}^{ctrl} + a_2\mathcal{L}^{data} + a_3\mathcal{L}^{mlm}
\end{equation}
\vspace{-0.1cm}
where $a_1$, $a_2$ and $a_3$ are the weights of the three losses.
In our implementation, we set $a_1=5$, $a_2=20$ and $a_3=1$ based on grid search on $PT_{valid}$.
PDBERT is pre-trained on 4 Nvidia RTX 3090 GPUs for 10 epochs with batch size 128.
We use the Adam optimizer~\cite{kingma2014adam} with an initial learning rate 1e-4 and apply the polynomial decay scheduler ($p=2$).
It takes approximately 72 hours to finish pre-training.
 \section{Evaluation}\label{sec:evaluation}
To demonstrate the effectiveness of PDBERT, we conduct both intrinsic and extrinsic evaluations.
For intrinsic evaluation, we directly use PDBERT to perform program dependence analysis for both partial and complete functions.
For extrinsic evaluation, we fine-tune and evaluate PDBERT on three function-level vulnerability analysis tasks, i.e., vulnerability detection, vulnerability classification, and vulnerability assessment.
Please note that since some baselines cannot handle partial functions, the extrinsic evaluation is conducted on complete functions.

\mysubsection{Intrinsic Evaluation}\label{sec:intrinsic}
\subsubsection{Motivation}
To understand to what extent PDBERT has learned the knowledge of program dependence, we apply PDBERT to perform CDP and DDP without further training or fine-tuning.

\subsubsection{Experimental Setup}\label{sec:in_exp_setup}
We conduct the intrinsic evaluation on $PT_{test}$, which is \textbf{never} used during pre-training.
PDBERT takes as input only a code snippet and predicts all the elements in its $G^c$ and $G^d$.
During evaluations, we do not assume that the types of code tokens are available and thus do not use the token-type-based masking strategy (cf. Section~\ref{sec:ddp}).
F1-score is used as the evaluation metric, which is calculated by flattening and concatenating the predicted graphs of all test samples.

We first evaluate the feasibility of PDBERT in analyzing partial code, as this is a unique benefit of PDBERT.
Specifically, we extract the first $K$ consecutive statements of each function in $PT_{test}$ to craft sub test sets.
$K$ is set to \{5,10,15,20,25,30\}.
For each sub test set, a unique $K$ is used and the functions with less than $K$ statements are ignored.
PDBERT takes as input each partial code snippet and predicts the program dependencies among its K statements.
To further understand the knowledge learned by PDBERT, we also evaluate PDBERT's effectiveness in analyzing complete functions.
Following CodeBERT, the max sequence length of PDBERT is set to 512.
Thus we filter out the complete functions with over 512 tokens in $PT_{test}$, and over 83\% of the functions in $PT_{test}$ are kept.
We partition the remaining functions into several non-overlapping subsets based on their Lines of Code (LOC) and apply PDBERT to them, aiming to show PDBERT's effectiveness across functions of varying lengths.
Besides, to demonstrate the throughput of PDBERT, we measure the average time cost of PDBERT to perform CDP and DDP for a complete function and compare it with Joern, the state-of-the-art program dependence analysis tool.
Specifically, PDBERT is deployed on one Nvidia RTX 3090 GPU to perform predictions with batch size 1.
Joern is run on two Intel Xeon 6226R CPUs with 64 cores in total using the default configuration.

\begin{table}[]
    \centering
    \caption{Intrinsic evaluation results on partial code in terms of F1-score.}
    \label{tab:partial_pdg_prediction_results}
\begin{tabular}{@{}ccccccc@{}}
    \toprule
    \textbf{Dependency} & \textbf{5} & \textbf{10} & \textbf{15} & \textbf{20} & \textbf{25} & \textbf{30} \\
    \midrule
    \textbf{Control} & 98.99 & 99.29 & 99.30 & 99.25 & 99.23 & 99.21 \\
    \textbf{Data} & 97.38 & 96.61 & 96.05 & 95.63 & 95.44 & 95.08 \\
    \textbf{Overall} & 97.69 & 97.51 & 97.29 & 97.07 & 97.00 & 96.84 \\
    \bottomrule
    \end{tabular}
\end{table} \begin{table}[]
    \centering
    \setlength\tabcolsep{3pt}
    \caption{Intrinsic evaluation results on complete functions in terms of F1-score.}
    \label{tab:pdg_prediction_results}
\begin{threeparttable}
    \begin{tabular}{@{}ccccc@{}}
    \toprule
    \textbf{Dependency} & \textbf{(0, 10]} & \textbf{(10, 20]} & \textbf{(20, 30]} & \textbf{(30, +$\infty$]} \\
    \midrule
    \textbf{Control} & 99.91 & 99.82 & 99.44 & 99.23 \\
    \textbf{Data} & 97.59 & 96.35 & 95.28 & 94.31 \\
    \textbf{Overall} & 98.07 & 97.56 & 96.93 & 96.46 \\
    \bottomrule
    \end{tabular}
    \begin{tablenotes}
        \footnotesize
        \item{*} (0, 10] denotes the subset where the LOC of each function is over 0 and no more than 10, and so forth.
    \end{tablenotes}
\end{threeparttable}
\end{table} 
\subsubsection{Results}
Table~\ref{tab:partial_pdg_prediction_results} shows the evaluation results of PDBERT on partial code in terms of F1-score.
The results presented in the ``Overall'' row are calculated on the combination of control and data dependencies.
We can see that PDBERT performs very well on CDP, achieving F1-scores of about 99\% for all Ks.
As for DDP, PDBERT achieves F1-scores of over 95\%, which is also impressive.
Note that for each code snippet, DDP instances are highly imbalanced.
Thus, DDP is more challenging than CDP.
The overall F1-scores of PDBERT are over 96\% for all Ks, indicating the feasibility and effectiveness of PDBERT in analyzing program dependencies for partial code.
We also notice that the control, data and overall F1-scores of PDBERT slightly decrease as $K$ increases.
This trend is attributed to the increased difficulty in analyzing long code snippets compared to shorter ones.
Please note that existing static analysis tools, e.g., Joern, lack the ability to automatically derive program dependencies in partial code, and thus are unsuitable for this scenario.

Table~\ref{tab:pdg_prediction_results} presents the performance of PDBERT on complete functions, which closely aligns with that on partial code.
Across all LOC ranges, PDBERT achieves control, data and overall F1-scores of over 99\%, 94\% and 96\%, respectively, demonstrating the effectiveness and potential utility of PDBERT in performing program dependence analysis on complete functions.
Similarly, there is a slight decrease in the control, data, and overall F1-scores as the LOC increases.

Regarding throughput, on average, Joern takes over 460ms to derive the PDG of a complete function, whereas PDBERT accomplishes the same task in just 19ms, making it 23 times faster than Joern.
Please note that this comparison may not be entirely fair, as PDBERT is deployed on a GPU while Joern is run on two CPUs.
The result only highlights the advantage of PDBERT in throughput and does not mean PDBERT can replace Joern in all use cases.
Nevertheless, it implies that PDBERT is more suitable than Joern for the use cases where some low levels of imprecision are tolerant and high throughput matters more.

\vspace{-0.05cm}
\begin{framed}
\vspace{-0.2cm}
To conclude, PDBERT has learned the knowledge of program dependence and can accurately extract program dependencies for both partial and complete functions with high throughput.
\vspace{-0.2cm}
\end{framed}

\mysubsection{Common Baselines and Variants for Extrinsic Evaluation}
\label{common baselines}
The extrinsic evaluation aims to investigate whether the pre-training of PDBERT is effective and can boost vulnerability analysis tasks.
To this end, we compare PDBERT with the pre-trained code models using other pre-training objectives and non-pre-trained models.
As a proof of concept, PDBERT is pre-trained based on an encoder-only Transformer model, because decoder-only models are sub-optimal for understanding tasks~\cite{guo2022unixcoder,devlin2019bert} and encoder-decoder models require much more data and time to converge~\cite{ahmad2021plbart,wang2021codet5}.
For example, Ahmad et al.~\cite{ahmad2021plbart} spent 2,208 GPU hours and used 727 million functions/documents to pre-train PLBART.
Following Ding et al.~\cite{ding2022learning}, to perform a fair comparison, we only consider encoder-only pre-trained models as baselines.

Specifically, for pre-trained code models, we use \textbf{CodeBERT}~\cite{feng2020codebert}, \textbf{GraphCodeBERT}~\cite{guo2021graphcodebert}, \textbf{VulBERTa}~\cite{hanzif2022vulberta} and \textbf{DISCO}~\cite{ding2022learning}
as baselines.
CodeBERT is widely used in various software engineering tasks~\cite{zhou2021finding,zhou2021assessing,le2022usea,huang2022prompt}, and PDBERT is initialized with the parameters of CodeBERT.
GraphCodeBERT might be considered somewhat similar to our DDP task (cf. Section~\ref{sec:pre_trained}).
Comparing PDBERT with GraphCodeBERT can help demonstrate the benefits of DDP and PDBERT.
VulBERTa~\cite{hanzif2022vulberta} is pre-trained by Hanzi et al. with their collected C/C++ functions (cf. Section~\ref{sec:dataset}) using MLM.
DISCO is the state-of-the-art encoder-only pre-trained code model, which uses MLM, node-type MLM (NT-MLM), and a contrastive learning objective for pre-training,
and is also evaluated on vulnerability detection.
Unfortunately, the pre-trained model of DISCO is not publicly available.
Therefore, we can only compare PDBERT with DISCO on vulnerability detection based on the results reported in its paper~\cite{ding2022learning}.
There are also other encoder-only pre-trained code models, including CuBERT~\cite{kanade2020learninga}, ContraCode~\cite{jain2021contrastive} and \textsc{Code-MVP}~\cite{wang2022codemvp}.
But they are pre-trained only on Python or JavaScript programs and hence are not suitable for C/C++ programs.
In addition, for most of them, the pre-trained models are not released.

For non-pre-trained models, we adopt a bidirectional LSTM (\textbf{Bi-LSTM})~\cite{graves2013speech} and a multi-layer \textbf{Transformer} model \cite{vaswani2017attention} as baselines.
They are frequently used as encoders to handle sequential inputs.
To eliminate the Out-of-Vocabulary (OoV) problem \cite{karampatsis2020bigcode}, their vocabularies and tokenizers are built using the BPE algorithm \cite{senrich2016neural}. 

We build four variants of PDBERT, namely \textbf{MLM}, \textbf{MLM+CDP}, \textbf{MLM+DDP} and \textbf{MLM+CDP+DDP}.
They are pre-trained using the same settings but with different pre-training objectives, as reflected by their names.
Comparing their performance can help understand the contributions of different objectives to different tasks.

\mysubsection{Vulnerability Detection}~\label{sec:vul_detection}
\subsubsection{Introduction}
Vulnerability detection is a fundamental problem in software security.
Following prior work~\cite{chakraborty2022reveal,ding2022learning}, this task aims to predict whether a function is vulnerable or not, i.e., function-level vulnerability detection.

\begin{table}[]
\setlength\tabcolsep{4pt}
\centering
\caption{The statistics of datasets.}
\label{tab:dataset_statistics}
\begin{threeparttable}
\begin{tabular}{@{}lccccccc@{}}
    \toprule
    \multirow{2}{*}{\textbf{Dataset}} &
    \multirow{2}{*}{\textbf{\#Func}} &
    \multirow{2}{*}{\textbf{\#Vul}} &
    \multirow{2}{*}{\textbf{\#Non-Vul}} &
    \multicolumn{2}{c}{\textbf{LOC}} &
    \multicolumn{2}{c}{\textbf{CC}} \\
    \cmidrule(l){5-6}
    \cmidrule(l){7-8}
    & & & & \textbf{\scriptsize{Mean}} & \textbf{\scriptsize{Median}} & \textbf{\scriptsize{Mean}} & \textbf{\scriptsize{Median}} \\
    \midrule
\textbf{ReVeal}       & 22.6K & 2.2K & 20.4K & 37.2 & 15 & 7.1 & 3 \\
    \textbf{Big-Vul}      & 188.6K & 10.9K & 177.7K & 25.0 & 12 & 5.9 & 3 \\
    \textbf{Devign}       & 27.2K & 12.4K & 14.8K & 51.8 & 26 & 11.2 & 5 \\
    \textbf{VC}           & 7.6K & 7.6K & N/A & 80.5 & 32 & 16.7 & 6 \\
    \textbf{VA}           & 9.9K & 9.9K & N/A & 79.1 & 32 & 16.3 & 6 \\
    \bottomrule
\end{tabular}
\begin{tablenotes}
    \footnotesize
\item{*} \#Func, \#Vul and \#Non-Vul refer to the numbers of all, vulnerable, and non-vulnerable functions. CC denotes cyclomatic complexity. VC and VA refer to the datasets used for vulnerability classification and vulnerability assessment, respectively.
\end{tablenotes}
\end{threeparttable}
\end{table} 
\subsubsection{Experimental Setup}
\label{vulnerability detection experimental setting}
We evaluate PDBERT on three widely-used C/C++ function-level vulnerability detection datasets, i.e., ReVeal~\cite{chakraborty2022reveal}, Big-Vul~\cite{fan2020dataset} and Devign~\cite{zhou2019devign}. Their statistics are presented in Table~\ref{tab:dataset_statistics}.

\textbf{ReVeal} is released by the ReVeal paper~\cite{chakraborty2022reveal}, and is collected from Linux Debian Kernel and Chromium.
This dataset is imbalanced and close to real-world scenarios.
We randomly split 70\%/10\%/20\% of the dataset for training, validation and testing, respectively, following prior work~\cite{chakraborty2022reveal,ding2022learning}.
Considering the data imbalance and following prior work~\cite{chakraborty2022reveal,ding2022learning}, we use F1-score as the evaluation metric.

\textbf{Big-Vul} is released by Fan et al. \cite{fan2020dataset}, and is collected from 348 Github projects.
Considering the large scale of this dataset and following prior work~\cite{li2021ivdetect}, we randomly split it into training, validation and testing sets by 80\%/10\%/10\%.
Since this dataset is highly imbalanced, we also use F1-score as the evaluation metric.

\textbf{Devign} is released by the Devign paper \cite{zhou2019devign}, containing 27.2K functions collected from FFmpeg and Qemu.
This dataset is balanced and less realistic than ReVeal and Big-Vul.
We use this dataset because it is included in the frequently used CodeXGLUE benchmark~\cite{lu2021codexglue}.
We also use the train/valid/test splits from CodeXGLUE.
Following the design of CodeXGLUE and prior work~\cite{ding2022learning}, we use Accuracy as the evaluation metric.

For PDBERT and the common baselines introduced in Section~\ref{common baselines}, a new MLP is appended to each model to perform prediction.
Apart from the common baselines, we also compare PDBERT with the state-of-the-art non-pre-trained models that are specially designed for vulnerability detection, 
including VulDeepecker \cite{li2018vuldeepecker}, Devign \cite{zhou2019devign}, SySeVR~\cite{li2021sysevr}, and ReVeal \cite{chakraborty2022reveal}.
For Devign, its implementation is not released, so we use the implementation and settings provided by the authors of ReVeal~\cite{chakraborty2022reveal}.
For other baselines, we use their official implementation and settings to conduct experiments.

\begin{table}[!t]
    \centering
    \caption{Evaluation results on vulnerability detection.}
    \label{tab:vul_detection_baselines}
\begin{threeparttable}
    \begin{tabular}{@{}lccc@{}}
    \toprule
    \multirow{4}{*}{\textbf{Model}} & \multicolumn{3}{c}{\textbf{Dataset}}                 \\ \cmidrule(l){2-4} 
                                & \begin{tabular}[c]{@{}c@{}} \textbf{ReVeal} \\ (F1-score)\end{tabular} & \begin{tabular}[c]{@{}c@{}}\textbf{Big-Vul} \\ (F1-score)\end{tabular} & \begin{tabular}[c]{@{}c@{}}\textbf{Devign}\\ (Accuracy)\end{tabular} \\ \midrule
    Bi-LSTM                         & 34.25           & 33.11           & 61.24            \\
    Transformer                     & 40.91           & 34.90           & 60.51            \\ 
VulDeePecker                    & 29.03           & 13.07            & 45.68            \\
Devign                          & 26.43           & 13.77           & 52.27               \\
    SySeVR                          & 33.73           & 24.80           & 48.87 \\
    ReVeal                          & 32.24           & 23.93                         & 54.55\\
\midrule
    CodeBERT                        & 44.27           & 54.48           & 62.08            \\
    GraphCodeBERT                   & 45.03           & 54.06           & 64.02            \\
VulBERTa & 44.19    & 36.75  & 62.88\\
    DISCO\dag           & 46.4*               & -         & 63.8 \\ 
\midrule
    \textbf{PDBERT} \\
    MLM                             & 44.65           & 55.99           & 65.52 \\
    MLM+CDP                         & 45.93           & 56.28           & 66.29            \\
    MLM+DDP                         & 47.42           & 58.51           & 65.85                \\
    MLM+CDP+DDP                     & \textbf{48.38}  & \textbf{59.41}  & \textbf{67.61}   \\ \bottomrule
    \end{tabular}
    \begin{tablenotes}
\scriptsize
        \item{\dag} The pre-trained model of DISCO is not publicly available. Its results are copied from the DISCO paper~\cite{ding2022learning}.
        \item{*} Since the ReVeal dataset does not provide official splits, the test set used by DISCO can be different from other approaches. 
\end{tablenotes}
\end{threeparttable}
\end{table} 

\subsubsection{Results}
Table~\ref{tab:vul_detection_baselines} presents our evaluation results on this task. 
We can see that the pre-trained models outperform all the non-pre-trained models on the three datasets by substantial margins, confirming the effectiveness of pre-training.
It is a little surprising that Bi-LSTM and Transformer perform better than the other non-pre-trained models.
One possible reason is that they both use BPE tokenizers, which have been shown to be effective in modeling code by prior work~\cite{karampatsis2020bigcode,thongtanunam2022autotransform}.
Ding et al.~\cite{ding2022learning} also reported similar results.
On the ReVeal, Big-Vul and Devign datasets, PDBERT improves the best-performing baselines, i.e., DISCO, CodeBERT, and GraphCodeBERT, by 4.3\%, 9.0\% and 5.6\%, respectively, in F1-score or Accuracy.
It is worth mentioning that among the baselines, the best-performing baseline on one dataset usually does not perform best on another.
Nevertheless, PDBERT consistently performs better than all the baselines.

For the variants of PDBERT, 
MLM+CDP+DDP outperforms MLM on the three datasets by substantial margins, highlighting the effectiveness of our pre-training objectives.
MLM+CDP+DDP performs better than MLM+CDP and MLM+DDP, indicating that CDP and DDP are both helpful.
In addition, MLM+DDP outperforms GraphCodeBERT, indicating the benefits of DDP.

\vspace{-0.05cm}
\begin{framed}
\vspace{-0.2cm}
In summary, PDBERT is more effective in vulnerability detection than the baselines on the three datasets. Both CDP and DDP contribute to PDBERT's effectiveness.
\vspace{-0.2cm}
\end{framed}

\mysubsection{Vulnerability Classification}

\subsubsection{Introduction}
After a vulnerability is detected, it will usually be labeled with a category to help practitioners understand its root cause, impact and possible mitigation~\cite{zou2019mu}.
However, this labeling process requires experts to manually analyze the vulnerability.
This task aims to automate this process by automatically classifying a vulnerable function based on its source code.
Specifically, we focus on the Common Weakness Enumeration (CWE) categories, which are adopted by many well-known software vulnerability databases, such as NVD~\cite{nvd} and VulDB~\cite{vuldb}, for vulnerability classification.
This task is a multi-class classification task.

\subsubsection{Experimental Setup}
\label{cwe classification experimental setting}
We construct a dataset for this task based on the Big-Vul dataset~\cite{fan2020dataset}.
Besides source code, the Big-Vul dataset also collects other vulnerability-related information, such as the CWE category and the Common Vulnerability Scoring System (CVSS) scores, for each vulnerable function.
To construct our dataset, we remove all the non-vulnerable functions.
We also filter out the vulnerable functions belonging to the CWE categories with less than 100 (about 1\%) instances, to avoid the absence of some CWE categories after data splitting.
The resulting dataset contains 7.6K vulnerable functions in 14 CWE categories.
We randomly split 80\%/10\%/10\% of this dataset for training/validation/testing.

Due to the long-tailed distribution of CWE categories, we use three metrics, i.e., Macro F1, Weighted F1 and the multi-class version of Matthews Correlation Coefficient (MCC)~\cite{gorodkin2004comparing}, for evaluation.
These metrics are also used by other vulnerability-related studies~\cite{han2017learning,le2021deepcva}.
Macro F1 is the unweighted mean of the F1-scores of all categories, whereas Weighted F1 considers weighted mean. 
MCC measures the differences between actual values and predicted values.
Note that MCC ranges from -1 to 1 and is not directly proportional to F1-score.
A new MLP is used by PDBERT and each of the common baselines, respectively, as the classifier.
To the best of our knowledge, existing approaches for CWE category prediction either are based on vulnerability descriptions or require inter-procedural analysis.
Considering that this task only takes as input vulnerable functions, there is no suitable task-specific baseline.

\begin{table}[!t]
    \centering
    \caption{Evaluation results on vulnerability classification.}
    \label{tab:cwe_classification_baselines}
\begin{tabular}{@{}lccc@{}}
    \toprule
    \textbf{Model} & \textbf{Macro F1(\%)} & \textbf{Weighted F1(\%)} & \textbf{MCC}    \\ \midrule
    Bi-LSTM        & 26.48             & 39.45                & 0.3001          \\
    Transformer    & 35.35             & 46.34                & 0.3737          \\ \midrule
    CodeBERT       & 51.91             & 57.37                & 0.5030          \\
    GraphCodeBERT  & 54.74             & 59.82                & 0.5331          \\
    VulBERTa       & 50.11             & 57.36                & 0.5035          \\
    \midrule
\textbf{PDBERT}  & & & \\
    MLM               & 54.49                & 60.42    & 0.5372 \\
    MLM+CDP           & 56.51                & 59.93    & 0.5325 \\
    MLM+DDP           & 56.93                & 62.07    & 0.5614 \\
    MLM+CDP+DDP       & \textbf{57.96}    & \textbf{62.60}       & \textbf{0.5644} \\ \bottomrule
    \end{tabular}
\end{table} 

\subsubsection{Results}
Table~\ref{tab:cwe_classification_baselines} presents the evaluation results on this task.
We can see that the pre-trained models outperform all the non-pre-trained baselines, i.e., Bi-LSTM and Transformer, by large margins.
For pre-trained models, VulBERTa achieves similar performance to CodeBERT.
PDBERT improves CodeBERT in Macro F1, Weighted F1 and MCC by 11.7\%, 9.1\% and 12.2\%, respectively.
PDBERT also outperforms GraphCodeBERT in the three metrics by 5.9\%, 4.6\% and 5.9\%, demonstrating its effectiveness in this task.

As for PDBERT's variants, 
MLM+CDP+DDP outperforms MLM in all the metrics by substantial margins, highlighting that our pre-training objectives are effective.
MLM+CDP+DDP improves MLM+CDP and MLM+DDP, which indicates that CDP and DDP are both beneficial.
Also, MLM+DDP outperforms GraphCodeBERT by substantial margins, confirming the benefits of DDP.

\vspace{-0.05cm}
\begin{framed}
\vspace{-0.2cm}
In summary, PDBERT can boost vulnerability classification and both CDP and DDP are effective and beneficial for this task.
\vspace{-0.2cm}
\end{framed}

\mysubsection{Vulnerability Assessment}
\subsubsection{Introduction}
Vulnerability assessment is a process that determines various characteristics of vulnerabilities and helps practitioners prioritize the remediation of critical vulnerabilities~\cite{le2021deepcva,le2023survey}.
CVSS is a commonly used expert-based vulnerability assessment framework.
It defines a series of metrics, i.e., CVSS metrics, to measure the severity of a vulnerability relative to other vulnerabilities.
However, quantifying these metrics for a new vulnerability requires manual efforts of security experts and there is usually a delay in such manual process~\cite{le2021deepcva}.
To this end, the vulnerability assessment task aims to automatically assess the CVSS metrics for vulnerabilities.
In this work, we focus on function-level vulnerability assessment, which takes as input a vulnerable function and outputs the values of CVSS metrics for it.
Specifically, this task targets four metrics that are important and could be inferred based on source code, including Availability, Confidentiality, Integrity, and Access Complexity.

\subsubsection{Experimental Setting}
We also construct a dataset for this task based on the Big-Vul dataset \cite{fan2020dataset}.
Specifically, we remove the vulnerable functions with invalid CVSS scores (e.g. ``???''), resulting in a dataset with 9.9K vulnerable functions and their CVSS scores.
We randomly split 80\%/10\%/10\% of them for training/validation/test.
We use the state-of-the-art vulnerability assessment approach named DeepCVA~\cite{le2021deepcva} as the task-specific baseline.
DeepCVA uses Convolutional Neural Networks (CNN) to extract features from vulnerable code and predict multiple CVSS metrics in parallel through multi-task learning.
To perform a fair comparison, we reuse the official implementation of DeepCVA, and train and evaluate DeepCVA on our dataset.
For each CVSS metric, we append a new MLP to PDBERT and each of the common baselines, respectively, as the classifier.
Following DeepCVA, we use two evaluation metrics, i.e., Macro F1 and the multi-class version of MCC.

\begin{table*}[!t]
  \centering
  \caption{Evaluation results on vulnerability assessment.}
\label{tab:vul_assessment_baselines}
  \begin{threeparttable}
  \begin{tabular}{@{}lcccccccccc@{}}
  \toprule
  \multirow{2}{*}{\textbf{Model}} &
    \multicolumn{2}{c}{\textbf{Access Complexity}} &
    \multicolumn{2}{c}{\textbf{Availability}} &
    \multicolumn{2}{c}{\textbf{Integrity}} &
    \multicolumn{2}{c}{\textbf{Confidentiality}} &
    \multicolumn{2}{c}{\textbf{Mean}} \\ \cmidrule(l){2-11} 
   &
    \textbf{\scriptsize{Macro F1(\%)}} &
    \textbf{\scriptsize{MCC}} &
    \textbf{\scriptsize{Macro F1(\%)}} &
    \textbf{\scriptsize{MCC}} &
    \textbf{\scriptsize{Macro F1(\%)}} &
    \textbf{\scriptsize{MCC}} &
    \textbf{\scriptsize{Macro F1(\%)}} &
    \textbf{\scriptsize{MCC}} &
    \textbf{\scriptsize{Macro F1(\%)}} &
    \textbf{\scriptsize{MCC}} \\ \midrule
  Bi-LSTM       & 59.20 & 0.4302 & 66.07 & 0.5423 & 72.26 & 0.5802 & 71.29 & 0.5589 & 67.21 & 0.5279 \\
  Transformer   & 47.33 & 0.3590 & 63.52 & 0.5115 & 71.33 & 0.5595 & 69.56 & 0.5256 & 62.93 & 0.4889 \\
DeepCVA       & 73.06 & 0.6093 & 74.50 & 0.6579 & 75.64 & 0.6394 & 75.30 & 0.6290 & 74.63 & 0.6337 \\ \midrule
  CodeBERT      & 69.69 & 0.6054 & 76.98 & 0.6908 & 77.59 & 0.6580 & 76.24 & 0.6368 & 75.13 & 0.6477 \\
  GraphCodeBERT & 74.78 & 0.6484 & 77.21 & 0.6895 & 76.72 & 0.6490 & 75.94 & 0.6305 & 76.16 & 0.6544 \\
VulBERTa & 68.97 & 0.6062 & 75.11 & 0.6615 & 77.46 & 0.6621 & 75.36 & 0.6184 & 74.23 & 0.6371 \\
  \midrule
  \textbf{PDBERT} \\
  MLM & 
  75.08 &
  0.6639 &
  75.64 &
  0.6782 &
  78.15 &
  0.6731 &
  77.46 &
  0.6482 &
  76.58 &
  0.6658 \\
  MLM+CDP &
  \textbf{79.75} &
  \textbf{0.7300} &
  \textbf{78.31} &
  \textbf{0.6998} &
  78.85 &
  0.6851 &
  \underline{77.48} &
  \textbf{0.6605} &
  \textbf{78.60} &
  \textbf{0.6939} \\
  MLM+DDP &
  76.58 &
  0.6655 &
  77.97 &
  \underline{0.6993} &
  \underline{79.37} &
  \textbf{0.6925} &
  77.26 &
  0.6474 &
  77.59 &
  0.6724 \\
  MLM+CDP+DDP & 
  \underline{78.06} &
  \underline{0.6682} &
  \underline{78.15} &
  0.6965 &
  \textbf{79.38} &
  \underline{0.6884} &
  \textbf{77.52} &
  \underline{0.6519} &
  \underline{78.28} &
  \underline{0.6763} \\ \bottomrule
  \end{tabular}
  \begin{tablenotes}
    \footnotesize
\item The best results are bold, and the second-best results are underlined.
\end{tablenotes}
\end{threeparttable}
\end{table*} 
\subsubsection{Results}
Table~\ref{tab:vul_assessment_baselines} presents the evaluation results on the vulnerability assessment task.
On average, the pre-trained models perform better or at least no worse than the best non-pre-trained model, i.e., DeepCVA, demonstrating the effectiveness of pre-training.
Note that assessing CVSS metrics often requires more context and project-specific information, e.g., the system configuration required to exploit the vulnerability, than detecting or classifying vulnerabilities.
Such information can hardly be found in the vulnerable function.
Therefore, the performance improvements that can be achieved by improving code understanding are limited.
Nevertheless, PDBERT outperforms the best-performing baseline, i.e., GraphCodeBERT, on each CVSS metric.
On average, MLM+CDP+DDP improves GraphCodeBERT in Macro F1 and MCC by 2.8\% and 3.3\%.
The performance improvements of the best-performing variant, i.e., MLM+CDP, over GraphCodeBERT in the two metrics are 3.2\% and 6.0\%, respectively.
These results indicate that our pre-training technique is effective in assessing vulnerabilities.

For the variants of PDBERT, 
MLM+CDP+DDP outperforms MLM on all the CVSS metrics and on average, highlighting the contribution of our pre-training objectives.
The performance of MLM+CDP and MLM+CDP+DDP is very close when assessing Availability, Integrity, and Confidentiality.
However, MLM+CDP performs better than MLM+CDP+DDP on Access Complexity.
One possible reason is that Access Complexity cares more about the conditions or privileges required to execute the vulnerable statements, which are highly related to control dependencies.
This shows that for vulnerability assessment, DDP does not provide significant benefits.
Besides, MLM+DDP still outperforms GraphCodeBERT, indicating the effectiveness of DDP.

\vspace{-0.05cm}
\begin{framed}
\vspace{-0.2cm}
In summary, PDBERT is effective in vulnerability assessment and CDP is more important than DDP for this task.
\vspace{-0.2cm}
\end{framed}
 \section{Discussion}\label{sec:disc}
This section discusses the limitations of our approach and the threats to validity of this work.

\vspace{-0.1cm}
\subsection{Limitations}
Due to the limitations of computation resources, following CodeBERT, we set the max sequence length of PDBERT as 512 in this work.
Consequently, when used for program dependence analysis, PDBERT cannot properly analyze code snippets with over 512 tokens.
However, as shown in Section~\ref{sec:in_exp_setup}, over 83\% of C/C++ functions in the test set constructed from open-source projects adhere to this length limit.
Based on our intrinsic evaluation, we argue that PDBERT is effective and efficient for analyzing most functions in practice.
When applied to downstream tasks, PDBERT truncates long code snippets before processing them.
Although this may negatively affect the performance, our extrinsic evaluation shows that PDBERT still effectively boosts downstream tasks.
On the other hand, this limitation comes from the Transformer architecture~\cite{vaswani2017attention}, not from our pre-training technique.
It can be mitigated by simply increasing the max sequence length of PDBERT, which requires more computation resources and advanced hardware.
Considering that our pre-training technique is orthogonal to the underlying model architecture, one can also adopt specialized model architectures for long code snippets, such as LongFormer~\cite{beltagy2020longformer} and LongCoder~\cite{guo2023longcoder}, to address this limitation.
Since this work focuses on pre-training techniques, we choose the most widely used Transformer architecture, follow the settings of prior work~\cite{feng2020codebert,guo2021graphcodebert}, and leave handling long code snippets as future work.

In addition, CDP focuses on statement-level control dependencies and does not handle the control flow within a single statement, e.g., ternary operators.
Future work may improve CDP by considering such control flow.

\vspace{-0.1cm}
\subsection{Threats to Validity}\label{sec:threats}
Generalization of PDBERT. 
PDBERT is pre-trained on C/C++ programs and may not be suitable for other programming languages.
However, our proposed pre-training objectives are language-agnostic.
Practitioners can pre-train their models with CDP and DDP for other or even multiple programming languages.

The limitations of using Joern.
To build ground truth for CDP and DDP, we use Joern to extract PDGs.
Therefore, PDBERT is expected to predict the program dependencies analyzed by Joern, and may share the same limitations as Joern.
However, Joern is the state-of-the-art source-code-based program dependence analysis tool.
It is of high accuracy and has been widely used for vulnerability-related tasks~\cite{zhou2019devign,chakraborty2022reveal,cao2022mvd}.
In addition, our pre-training technique can be regarded as training the model to learn the program dependence analysis knowledge of Joern, which can still be beneficial (as shown by our extrinsic evaluation).
Our intrinsic evaluation demonstrates the unique benefits of PDBERT in program dependence analysis and at least assesses how close PDBERT is to Joern in analyzing complete functions.
Therefore, we argue that the limitations of Joern do not affect the conclusions of this work.

Data Contamination.
It is possible that some functions in the pre-training dataset also appear in some fine-tuning datasets, as they are all collected from open-source repositories.
However, the pre-training dataset contains both vulnerable and benign functions, and the task-specific labels of each function are unknown during pre-training.
In addition, prior work~\cite{brown2020languagea} has shown that data contamination may have little impact on the performance of pre-trained models.
Therefore, we argue that this threat is limited.

  \section{Related Work}\label{sec:related_work}

\mysubsection{Pre-Trained Code Models}\label{sec:pre_trained}
Existing pre-trained code models can generally be divided into encoder-only~\cite{kanade2020learninga,feng2020codebert,jain2021contrastive,guo2021graphcodebert,ding2022learning,wang2022codemvp}, decoder-only~\cite{svyatkovskiy2020intellicodea,liu2020multitask,li2022competition} and encoder-decoder models~\cite{mastropaolo2021studyinga,ahmad2021plbart,jiang2021treebert,wang2021codet5,chakraborty2022natgena,zhang2022coditt5,li2022automating,niu2022sptcode,guo2022unixcoder}.
As a proof of concept, this work focuses on pre-training an encoder-only model for vulnerability analysis.
For existing encoder-only pre-trained models, some of them~\cite{feng2020codebert,kanade2020learninga} directly adopt the pre-training tasks designed for NL.
For example, CodeBERT~\cite{feng2020codebert} adopts MLM~\cite{devlin2019bert} and Replaced Token Detection (RTD)~\cite{clark2020electra}.
These models can capture the naturalness of code.
Recently, some pre-training techniques~\cite{wang2022codemvp,ding2022learning} are proposed to help models learn the syntactic structure of code.
For example, DISCO~\cite{ding2022learning} adopts a pre-training objective to predict the masked AST types in the input.
Some prior work~\cite{jain2021contrastive,bui2021selfsupervised,ding2022learning,wang2022codemvp} also leverages contrastive learning (CL) objectives to help models learn high-level functional similarities between programs.
Different from them, our pre-trained objectives aim to help models learn to capture the semantic structure of code.

To the best of our knowledge, only two pre-trained code models, i.e., \textsc{Code-MVP}~\cite{wang2022codemvp} and GraphCodeBERT~\cite{guo2021graphcodebert}, consider the semantic structure of code (e.g., control- and data-flow information).
\textsc{Code-MVP} explicitly takes control flow graph (CFG) as input during pre-training and target learning representation \textbf{from} CFG.
GraphCodeBERT is pre-trained by predicting a few masked data flow edges with other unmasked ones and a variable sequence as input.
During inference, the variable sequence and all data flow edges of a program need to be extracted as input.
This implies that GraphCodeBERT also targets learning representation \textbf{from} data flow, as mentioned in its paper~\cite{guo2021graphcodebert}.
Compared to them, first, PDBERT considers control and data dependencies simultaneously, thus it can learn more comprehensive knowledge about program dependence.
More importantly, PDBERT targets learning the representation that \textbf{encodes} the semantic structure of code and is designed to \textbf{``absorb'' the knowledge required for end-to-end program dependence analysis}, which has not been investigated.
As discussed in Section~\ref{sec:intro} and demonstrated in Section~\ref{sec:evaluation}, this design brings several unique and significant benefits:
(1) PDBERT only takes source code as input and can properly process ``unparsable'' code.
(2) The knowledge learned by PDBERT is more general and can better boost downstream tasks. 
(3) PDBERT can directly be used to analyze statement-level control dependencies and token-level data dependencies, which to the best of our knowledge, cannot be achieved by existing neural models.
Therefore, we believe PDBERT provides significant contributions compared to existing pre-trained code models.

\mysubsection{Deep Learning for Vulnerability Analysis}
Many deep-learning-based approaches have been proposed for vulnerability analysis~\cite{li2018vuldeepecker,zhou2019devign,zou2019mu,chakraborty2022reveal,le2021deepcva}.
For \textbf{vulnerability detection}, the state-of-the-art approaches first leverage static analysis tools, e.g., Joern, to extract PDGs, and represent programs in different forms, such as code gadgets~\cite{li2018vuldeepecker}, syntax-based, semantics-based, and vector representations~\cite{li2021sysevr}, or graph-based representations~\cite{zhou2019devign,chakraborty2022reveal,li2021ivdetect}, based on their PDGs.
Then, they leverage different neural models, such as Bi-LSTM~\cite{li2018vuldeepecker}, CNN~\cite{zhou2019devign,li2021sysevr}, and GNN~\cite{zhou2019devign,li2021ivdetect,chakraborty2022reveal}, to extract the feature vector of each input program for vulnerability detection.
Compared to these approaches, PDBERT does not require program dependencies as input and can handle partial code, providing unique advantages.
Also, these approaches are trained from scratch to learn how to use program dependencies on a specific task, while PDBERT is pre-trained to learn the knowledge of program dependence and can be easily applied to multiple vulnerability analysis tasks.
For \textbf{vulnerability classification}, most existing studies focus on predicting CWE categories based on expert-curated vulnerability descriptions~\cite{ruohonen2018validation,aota2020automation}.
$\mu$VulDeePecker~\cite{zou2019mu} is the only work that predicts CWE categories based on source code, but it uses inter-procedural analysis, while our work focuses on function-level vulnerabilities.
For \textbf{vulnerability assessment}, most prior work predicts the CVSS metrics based on vulnerability descriptions instead of source code~\cite{spanos2018multitarget,le2019automated}.
DeepCVA~\cite{le2021deepcva} leverages CNN and multi-task learning for commit-level vulnerability assessment.
Le et al.~\cite{le2022usea} proposed to predict the CVSS metrics based on vulnerable statements, which require significant manual efforts to identify or require vulnerability fixes to be known.

 \section{Conclusion and Future Work}\label{sec:conclusion}
In this work, we propose two novel pre-training objectives, i.e., CDP and DDP, to help incorporate the knowledge of end-to-end program dependence analysis into neural models and boost vulnerability analysis.
CDP and DDP aim to predict the statement-level control dependencies and the token-level data dependencies in a program only based on its source code.
As a proof of concept, we build a pre-trained model named PDBERT with CDP and DDP, which achieves F1-scores of over 99\% and 94\% for predicting control and data dependencies, respectively, in partial and complete functions.
We also fine-tune and evaluate PDBERT on three vulnerability analysis tasks, i.e., vulnerability detection, vulnerability classification, and vulnerability assessment.
Experimental results indicate PDBERT benefits from CDP and DDP and it is more effective than the state-of-the-art baselines on all these tasks.

In the future, we would like to investigate how to pre-train encoder-decoder models with CDP and DDP. We plan to investigate the effectiveness of PDBERT on more downstream tasks and pre-train PDBERT with a multilingual corpus.
In addition, for the downstream tasks where static analyzers can successfully extract the program dependencies in the input code, it would be interesting to investigate whether combining our pre-training technique with explicitly provided program dependencies can further improve performance.

\begin{acks}
This research/project is supported by the National Natural Science Foundation of China (No. 62202420) and the Fundamental Research Funds for the Central Universities (No. 226-2022-00064). Zhongxin Liu gratefully acknowledges the support of Zhejiang University Education Foundation Qizhen Scholar Foundation.
\end{acks}

\balance
\bibliographystyle{ACM-Reference-Format}
\bibliography{references}

\end{document}